\documentclass[twocolumn,showpacs,prl]{revtex4}
\usepackage{graphicx}%
\usepackage{dcolumn}
\usepackage{amsmath}

\begin{document}

\preprint{HEP/123-qed}

\title{Dependence of spin susceptibility of a two-dimensional electron system on the valley degree of freedom}
\author{Y. P. Shkolnikov, K. Vakili, E. P. De Poortere, and M. Shayegan}
\affiliation{Department of Electrical Engineering, Princeton University,
Princeton, New Jersey 08544}

\date{\today}

\begin{abstract}
We report measurements of the spin susceptibility, $\chi\propto g_v g^*m^*$, in
an AlAs two-dimensional electron system where, via the application of in-plane
stress, we transfer electrons from one conduction-band valley to another ($g_v$
is the valley degeneracy, and $m^*$ and $g^*$ are the electron effective mass
and g-factor). At a given density, when the two valleys are equally populated
($g_v=2$), the measured $g^*m^*$ is smaller than when only one valley is
occupied ($g_v=1$). This observation counters the common assumption that a
two-valley two-dimensional system is effectively more dilute than a
single-valley system because of its smaller Fermi energy.
\end{abstract}


\maketitle

An unsettled question regarding the physics of a dilute,
two-dimensional electron system (2DES) concerns the behavior of
its spin susceptibility, $\chi=(\mu_B m_0 /2 \pi
\hbar^2)(g_vg^*m^*)$, where $\mu_B$ is the Bohr magneton and
$\hbar$ the Planck constant, $g_v$ is the valley degeneracy, and
$m^*$ and $g^*$ are the electron effective mass (in units of free
electron mass $m_0$) and g-factor, respectively. As the 2DES
density is lowered, the Coulomb energy of the system dominates
over its kinetic energy. A measure of the diluteness is the $r_s$
parameter, the inter-electron spacing measured in units of the
effective Bohr radius, or equivalently, the Coulomb energy
measured in units of the kinetic (Fermi) energy. Theory
\cite{Attaccalite2002} predicts that $g^*m^*$ increases with
increasing $r_s$ and eventually diverges above a critical $r_s$,
where the 2D electrons attain a ferromagnetic ground state.
Experimentally, although there is yet no consensus as to the
divergence of $g^*m^*$, most measurements in dilute 2DESs show an
increasing $g^*m^*$ with $r_s$
\cite{Okamoto1998,Shashkin2001,Pudalov2002,Zhu2003}. In this
Letter, we report measurements of $g^*m^*$ in an AlAs 2DES where
we change the valley occupation via the application of uniaxial
in-plane stress. The data add an interesting twist to this problem
as they reveal that, at a fixed density, $g^*m^*$ depends on the
valley degree of freedom in an unexpected manner: compared to
$g^*m^*$ for a single-valley system, $g^*m^*$ is smaller when two
valleys are occupied. This observation appears to be at odds with
the widely made assumption that a two-valley system is effectively
more dilute than its single-valley counterpart because of its
smaller Fermi energy \cite{Abrahams2001,Shashkin2003,Dharma2004}.

Our samples contain 2D electrons confined to a modulation doped, 11 nm-wide,
AlAs quantum well grown on a GaAs (001) substrate using molecular beam epitaxy
\cite{Etienne2002}. The quantum well is flanked by Al$_{0.4}$Ga$_{0.6}$As
barrier layers. Using metal electrodes (gates) deposited on the front and back
sides of the sample and illumination, we can vary the 2DES density, $n$,
between $2.9\times10^{15}$ and $7.4\times10^{15}$ m$^{-2}$. In this density
range, the low temperature electron mobility is about 20 m$^2$/Vs. The
resistance is measured using a lock-in amplifier at 0.3K on a Hall bar sample
aligned with the [100] crystal direction. The sample is mounted on a tilting
stage so that its orientation with respect to the applied magnetic field can be
varied. We studied three samples from two different wafers. Here we present
data from one sample; the measurements on other samples corroborate the
reported results.

We first show how valley occupancy can be tuned in our samples. In
bulk AlAs, conduction band minima (or valleys) occur at the six
equivalent X-points of the Brillouin zone. The constant energy
surface consists of six half-ellipsoids (three full ellipsoids in
the first Brillouin zone), with their major axes along one of the
$<$100$>$ directions; here we designate these valleys by the
direction of their major axes. These valleys are highly anistropic
with longitudinal and transverse effective mass $\simeq1.0$ and
$0.2$, respectively. In an AlAs quantum well with width
$\widetilde{>} 5$ nm, grown on a GaAs (001) substrate, biaxial
compression due to the lattice mismatch between AlAs and GaAs
raises the energy of the [001] valley so that only the [100] and
[010] valleys, with their major axes lying in the plane, are
occupied \cite{Maezawa1992}. This is the case for our samples. In
our experiments, we apply additional, uniaxial compression along
the [100] direction to transfer electrons from the [010] valley to
the [100] valley while the total 2D electron density remains
constant. We measure and compare $g^*m^*$ for the two cases where
all the electrons are either in the [100] valley ($g_v=1$) or are
distributed equally between the [100] and [010] valleys ($g_v=2$).
Note that in our work, $g_v=2$ refers strictly to an equal
electron concentration in the two valleys.

We apply stress to the sample by gluing it on the side of a commercial
piezoelectric (piezo) stack actuator with the sample's [100] crystal direction
aligned to the poling direction of the piezo \cite{shayegan2003}. Under a
positive applied voltage bias, the piezo stack expands along the poling
direction and shrinks in the perpendicular directions. This deformation and the
resultant stress are transmitted to the sample through the glue. Using this
technique, we can achieve a strain range of $4.7\times10^{-4}$ which, for
$n<4.5\times10^{15}$ m$^{-2}$, is large enough to transfer all the electrons
into a single valley \cite{epsilonxy}. Since the maximum valley splitting (3
meV) is much smaller than the $\sim$ 150 meV conduction band offset between the
AlAs well and the barriers, we expect a negligible strain-induced change
($<$1\%) in $n$, consistent with our transport data in perpendicular magnetic
fields. Using a calibrated, metal strain gauge glued to the opposite side of
the piezo, we monitor the applied strain with a relative accuracy of 5\%.

In our experiments, we carefully analyze the frequency composition
of the Shubnikov-de Hass oscillations as a function of strain.
From such data, we determine the electron densities in the
valleys. Additionally, we monitor the dependence of sample's
resistance on strain (piezoresistance) at zero and finite magnetic
fields.  The saturation of the piezoresistance at large (negative)
strains signals the onset of the [010] valley depopulation; this
depopulation is confirmed by the Shubnikov-de Haas data. From such
measurements, we know, for example, that for the data shown in
Fig. 1, electrons occupy the [100] and [010] valleys equally in
the second trace from bottom (bold), and that for the top two
traces (bold) which were taken at strain values of $-4.3$ and
$-3.9\times10^{-4}$, only the [100] valley is occupied (note that
these two traces completely overlap in the entire field range).

\begin{figure}
\includegraphics[scale=.85]{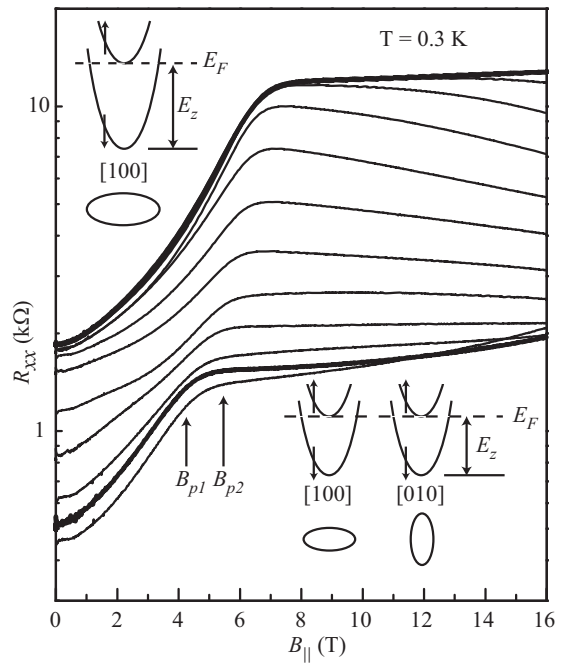}
\caption{Resistance of an AlAs 2DES at $n=2.85\times10^{15}$
m$^{-2}$ as a function of in-plane field applied along [100].
Strain increases from $-4.3\times10^{-4}$ to $3.9\times10^{-5}$
(top to bottom) in steps of $3.9\times10^{-5}$ (thirteen traces
are shown, the top two are completely overlapping). Negative
strain indicates a uniaxial compression along the [100] direction.
In the top two (bold) traces, only the [100] valley is occupied
($g_v=1$), while the second to the bottom trace (bold) corresponds
to the case where the 2D electrons are equally distributed between
the two [100] and [010] valleys ($g_v=2$). The position of the
pronounced kink seen in each trace is taken as the field $B_p$
above which the 2DES becomes fully spin polarized. The vertical
arrows $B_{p1}$ and $B_{p2}$ mark the possible range [$B_{p1}$,
$B_{p2}$] of $B_p$ for the lowest trace; we define $B_{p1}$ and
$B_{p2}$ as the fields at which the resistance deviates from its
exponential dependence on $B$ above and below the kink position.
The top and bottom insets schematically show the valley and
spin-subband occupations at $B_p$ for $g_v=1$ and $g_v=2$,
respectively.} \label{Bp}
\end{figure}

We employ two commonly used methods to measure $g^*m^*$. Both involve the
application of a magnetic field, $B$, but each determines $g^*m^*$ at a
different degree of spin polarization. In the first technique, we measure the
magneto-resistance (MR) of the sample as a function of $B$ applied strictly in
the plane of the 2DES (Fig. \ref{Bp}). With increasing $B$, the Zeeman
splitting $E_Z=g^*\mu_B B$ between the electrons with oppositely polarized
spins also increases and, at certain field ($B_p$), equals the Fermi energy,
$E_F$. Beyond $B_p$, the 2DES becomes fully spin-polarized. At $B_p$, we have:
\begin{equation}\label{B_kink}
    g^*m^*=(2 \pi \hbar^2/m_0 \mu_B)(n/g_v B_p).
\end{equation}
Because of the depopulation of one of the spin subbands at $B_p$, the electron
scattering rate changes, resulting in a kink in the MR trace
\cite{Okamoto1998}.

We show examples of such traces for $n=2.85\times10^{15}$ m$^{-2}$ in Fig.
\ref{Bp}. Each trace corresponds to a different amount of strain, between
$-4.3\times10^{-4}$ and $3.9\times10^{-5}$, applied along [100]. In all traces,
$B$ is also applied along [100]. A pronounced kink is observed in all the
traces. The position of the kink is the highest for the top two traces, where
all the electrons occupy the [100] valley. As the compression is decreased and
the electrons are transferred to the [010] valley, the kink moves to lower
fields.  The kink's field position is the smallest for the second trace from
bottom; for this trace the magnitude of strain is the smallest, and the two
[100] and [010] valleys have equal populations. The lowest trace corresponds to
positive (tensile) strain along [100]; we again have two unequally populated
valleys, but now the [010] valley has a larger population than the [100]
valley. For this trace the kink moves back to a slightly higher field.

Associating the kink position with the full spin polarization field $B_p$, the
above behavior can be qualitatively understood from Eq. (1) and the schematic
energy diagrams shown in Fig. \ref{Bp}. Given a fixed $n$, the Fermi energy of
the system is twice larger when all the electrons are in one valley ($g_v=1$)
compared to when they are distributed equally between the two valleys
($g_v=2$). If $g^*m^*$ remained constant, we would thus expect a twice smaller
$B_p$ in the case of $g_v=2$. This simple, non-interacting picture, however, is
not quantitatively consistent with the experimental data, which show a
reduction of only about 34\% in the kink position. According to Eq. (1), this
observation implies that $g^*m^*$ for $g_v=2$ is smaller than for $g_v=1$.

\begin{figure}
\includegraphics[scale=.85]{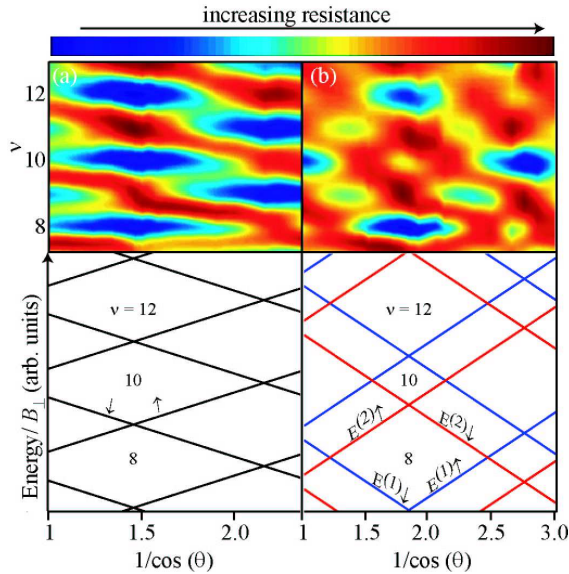}
\caption{(color) Angular dependence of the measured magneto-resistance (top
panels) and the associated energy fan diagrams (bottom panels) for (a) $g_v=1$
and (b) $g_v=2$ AlAs 2DES with $n=4.32\times 10^{15}$ m$^{-2}$. $E^{(1)}$ and
$E^{(2)}$ are the Landau energy levels for the [100] and [010] valleys,
respectively.} \label{tilt}
\end{figure}

Before discussing the above observation in detail, we present our measurements
of $g^*m^*$ using a second technique, the coincidence method \cite{Fang1968},
where we no longer restrict $B$ to the plane of the 2DES. With a non-zero
$B_{\perp}$ (out of plane component of $B$), Landau levels (LLs), with energy
separation equal to the cyclotron energy $E_c=\hbar eB_\perp/(m^*m_0)$, form.
Each LL is further split by $E_Z$, which is a function of total $B$. In a
typical measurement, we tilt $B$ by an angle $\theta$ from the 2DES normal,
such that the in-plane component of $B$ is along the [100] direction, and sweep
the magnetic field. At certain $\theta$ and $B$, the energy levels of
oppositely spin-polarized electrons coincide at the 2DES Fermi energy. If this
"coincidence" occurs when electrons occupy an integer number of LLs (integer
filling factor $\nu$), the MR minimum will rise \cite{Fang1968}. At such
coincidence, we have $g^*m^*=2\ N\ cos(\theta)$, where the integer $N$ can be
deduced from consecutive coincidences (in $\theta$) at a fixed $\nu$. We show
an example of such MR data for a case where $n=4.32\times10^{15}$ m$^{-2}$ and
all the electrons are in the [100] valley ($g_v = 1$) in the top panel of Fig.
\ref{tilt} (a). The corresponding energy level fan diagram is shown in the
bottom panel. At all odd $\nu>7$ (and similarly at all even $\nu>6$), $\theta$
of the coincidences are the same. This indicates that $g^*m^*$ is independent
of $B$ in this high filling range. From the data shown in Fig. \ref{tilt} (a),
we deduce $g^*m^*$ = 2.8.

The data shown in top panel of Fig. \ref{tilt} (b) were taken for the same $n$
as in the (a) panels, but at zero strain so that the [100] and [010] valleys
are equally occupied at $B=0$. When both valleys are occupied, the dependence
of MR on $\theta$ becomes more complicated, with coincidence angles forming a
"diamond" pattern with periodicity of four in $\nu$ \cite{Shkolnikov2002}. This
dependence of MR can be explained by an energy fan diagram (lower panel of Fig.
\ref{tilt} (b)) in which $E_v$ (energy splitting between valleys) and $E_c$
increase linearly with $B_{\perp}$, and $E_Z$ increases linearly with total $B$
and is the same for both valleys \cite{Shkolnikov2002}. By matching the values
of coincidence angles predicted by the energy fan diagram to the the
experimental data, we extract values for both $g^*m^*$ and $E_v$. The
coincidences at all odd $\nu>6$ occur at the same $\theta$ in data of Fig.
\ref{tilt}(b), implying that $g^*m^*$ is independent of $B$ for high fillings
in the $g_v=2$ case also. From the data of Fig. \ref{tilt} (b), we find
$g^*m^*=2.2$, again smaller than 2.8 that we find for the case of $g_v=1$ for
the same density.

\begin{figure}
\includegraphics[scale=.9]{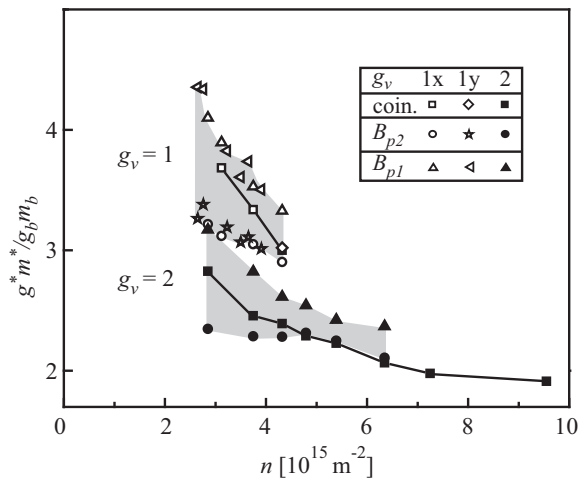}
\caption{The product $g^*m^*$ in units of $g_bm_b$ (product of band mass and
band g-factor) determined from both coincidence and $B_p$ measurements as a
function of $n$. Data are shown for when all electrons occupy the [100] valley
($g_v=1$, open symbols; orientation of the in-plane component of $B$ along
[100] and [010] is indicated by x and y, respectively), or are distributed
equally between [100] and [010] valleys ($g_v=2$, closed symbols). The ranges
of possible $g^*m^*$ from the parallel field measurements (i.e.,
[$B_{p1}$,$B_{p2}$]) are shown as shaded bands. Uncertainty of $g^*m^*$ deduced
from the coincidence measurement is less than 4\%.} \label{density}
\end{figure}

Figure \ref{density} captures the highlight of our study. It
summarizes the results of measurements for the cases where all the
electrons occupy the [100] valley or are distributed equally
between the [100] and [010] valleys. Plotted are the values of
$g^*m^*$ deduced from both the coincidence method, and the kink
positions in the parallel MR data and using Eq. (1) \cite{FN4}.
Since the two methods measure $g^*m^*$ at very different degrees
of spin polarizations - as low as 20\% in the coincidence method,
and 100\% in the $B_p$ method - the overlap of $g^*m^*$ from these
methods asserts that there is negligible non-linearity in spin
polarization, implying a nearly constant $g^*m^*$ as a function of
spin polarization \cite{FN2}. To check whether the values of
$g^*m^*$ are isotropic, we repeated both coincidence and $B_p$
measurements while applying the in-plane component of the magnetic
field along the [010] rather than [100] direction. The data shown
in Fig. \ref{density} include results from both field
orientations, evincing that $g^*m^*$ is indeed isotropic.

The results presented in Fig. \ref{density} are puzzling. In an
independent electron picture, $g^*m^*$ should not depend on the 2D
density or the valley degeneracy, clearly in disagreement with the
experimental data shown here. Electron-electron interaction, on
the other hand, is known to increase $g^*m^*$ as the density is
lowered and the system becomes more dilute \cite{Attaccalite2002}.
This trend is indeed seen in the data of Fig. \ref{density} for
both $g_v=1$ and $g_v=2$ \cite{FN1}. At any given density,
however, the measured $g^*m^*$ for the $g_v=1$ case is larger than
$g^*m^*$ for $g_v=2$. This is surprising since the $g_v=2$ system
is expected to be more "dilute" because of its smaller Fermi
energy.

In view of our results, it is useful to ask what the proper
definition of the $r_s$ parameter for a two-valley system is. This
parameter is commonly defined as the ratio of the Coloumb to
kinetic (Fermi) energy, $(e^2/4\pi\epsilon r)/E_F$, where $r$ is
the average inter-electron distance.  For a one-valley system,
$r_s$ is well defined, and is equal to $(1/a_B)(1/\pi n)^{1/2}$,
where $a_B$ is the effective Bohr radius. For a two-valley system,
there appears to be no clear consensus on the definition of $r_s$
\cite{Okamoto1998, Shashkin2003, Dharma2004}.  In all reported
definitions, however, it is assumed that, at a fixed density, the
two-valley system is equally or more dilute than the single-valley
system.  For example, in a model where intervalley interaction is
ignored, $r_s$ for the two-valley system is larger by a factor of
$\sqrt{2}$ (see for example, Ref. \cite{Dharma2004}). It is clear
from our data, that \textit{any} definition of $r_s$ that assumes
the two-valley system is equally or more dilute than the
one-valley system will be in apparent disagreement with our data.

Can our results be explained by some unusual properties of 2DESs
in AlAs? While we cannot rule out subtle or unknown phenomena, we
remark on three possibilities. First, $m^*$ can depend on valley
population if the energy dispersion is non-parabolic. Our
measurements \cite{shayegan2003} of the sample resistance as a
function of strain (valley population), while $n$ and $B_\perp$
(i.e., $\nu$) are kept constant, however, provide evidence against
this possibility: The resistance oscillates \textit{periodically}
with strain, implying that $m^*$ does not vary with valley
occupancy. Second, it is possible that when $g_v=2$, the lower
Fermi energy of the system leads to its being more susceptible to
disorder \cite{FN3}. While the role of disorder in modifying
$g^*m^*$ is not fully known, recent calculations \cite{Asgari2004}
indicate an increase in $m^*$ when disorder is present; opposite
to our observation of a smaller $g^*m^*$ for $g_v=2$. Third, the
Fermi contours in our AlAs 2DES are highly anisotropic with a
longitudinal to transverse effective mass ratio of approximately
5:1, implying anisotropic Bohr radius for electrons in a given
valley. A tantalizing possibility might be that the anomalous
behavior we observe is an indication of anisotropic interaction.

\begin{acknowledgments}
We thank the NSF for support, and E. Tutuc and R. Winkler and K. Karrai for
illuminating discussions.
\end{acknowledgments}


\begin{thebibliography}{19}
\bibitem[{eps()}]{epsilonxy}
\bibinfo{note}{Here we refer to the shear strain, defined as ($\Delta
  L_x/L_x-\Delta L_y/L_y$), where $L_x$ and $L_y$ are the sample dimensions
  along [100] and [010], respectively. The splitting between the $<$100$>$
  valleys is proportional to this strain.}
\bibitem[{FN3()}]{FN3}
\bibinfo{note}{As is evident from the data of Fig. 1, the mobility of the 2DES
  at $B_{||}=0$ is smaller when $g_v=1$. This is not because of larger
  disorder, but rather a consequence of the geometry of our measurement.
  Mobility of electrons is determined by the transport mass. When all the
  electrons are in the [100] valley, transport mass along [100] (the
  measurement direction) is just the longitudinal effective mass, whose larger
  value leads to the smaller observed mobility.}
\end{thebibliography}
\end{document}